\documentclass[draft]{agujournal2019}
\usepackage{url}
\usepackage{lineno}
\usepackage{soul}
\usepackage{amsmath,textcomp}

\journalname{Geophysical Research Letters}

\begin{document}

\title{Coexistence of two dune growth mechanisms in a landscape-scale experiment}

\authors{Ping L\"u\affil{1}, Cl\'ement Narteau\affil{2}, Zhibao Dong\affil{1}, Philippe Claudin\affil{3}, S\'ebastien Rodriguez\affil{2}, Zhishan An\affil{4}, Cyril Gadal\affil{2}, Sylvain Courrech du Pont\affil{5}}

\affiliation{1}{
School of Geography and Tourism, Shaanxi Normal University,
Xi'an, China.
\vskip 0.0cm
}
\affiliation{2}{
Universit\'e de Paris, Institut de physique du globe de Paris, CNRS,
Paris, France.
\vskip 0.0cm
}
\affiliation{3}{
Physique et M\'ecanique des Milieux H\'et\'erog\`enes,
CNRS, ESPCI PSL Research Univ, Sorbonne Univ, Universit\'e de Paris,
Paris, France.
\vskip 0.0cm
}
\affiliation{4}{
Northwest Institute of Eco-Environment and Resources,
Lanzhou, China.
\vskip 0.0cm
}
\affiliation{5}{
Laboratoire Mati\`ere et Syst\`eme Complexes, CNRS,
Universit\'e de Paris,
Paris, France.
\vskip 2cm
}

\correspondingauthor{Cl\'ement Narteau}{narteau@ipgp.fr}

{\vspace{1cm}\color{blue}\small  An edited version of this paper was published by AGU. Copyright 2020 American
Geophysical Union: Lü, P., Narteau, C., Dong, Z., Claudin, P., Rodriguez, S., An, Z., Gadal, C. \& Courrech du Pont, S. (2022). Coexistence of two dune growth mechanisms in a landscape‐scale experiment. Geophysical Research Letters, e2021GL097636.  https://doi.org/10.1029/2021GL097636}.


\begin{keypoints}
\item
Dunes of different shapes and orientations develop and coexist under the
same natural wind regime depending on sand availability.
\item
The dynamics of pattern coarsening selects dune aspect-ratio over short time.
\item
There is a minimum size for dune elongation on a nonerodible bed, below which
barchan or asymmetric barchan shapes are observed.
\end{keypoints}


\begin{abstract}
In landscape-scale experiments at the edge of the Gobi desert, we show that
various dune types develop simultaneously under natural wind conditions. Using
4 years of high-resolution topographic data, we demonstrate that, depending on
sand availability, the same wind regime can lead to two different dune
orientations, which reflect two independent dune growth mechanisms. As periodic
oblique dunes emerge from a sand bed and develop to 2 meters in height, we analyze
defect dynamics that drive the non-linear phase of pattern coarsening. Starting
from conical sand heaps deposited on gravels, we observe the transition from
dome to barchan and asymmetric barchan shapes. We identify a minimum size for
arm elongation and evaluate the contribution of wind reversals to its
longitudinal alignment. These experimental field observations support existing
theoretical models of dune dynamics boosting confidence in their applicability
for quantitative predictions of dune evolution under various wind regimes and
bed conditions.
\end{abstract}

\section*{Plain Language Summary}
Dune fields are characterized by the occurrence of both isolated dunes and periodic
bedforms of variable shape and orientations. However, there is no field evidence
whether these isolated and periodic dune patterns develop at the same time and from
the same growth mechanism. Here, by leveling neighboring parcels of a dune field,
we perform landscape-scale experiments with controlled initial and boundary
conditions to test the influence of sand availability on the formation and
timescales associated with the development of both types of patterns.
Starting from a flat sand bed, we observe the emergence of periodic dunes and
measure for more than 3 years how they grow as they interact with each other.
Over the same time period, by regularly feeding sand heaps deposited nearby on a
non-erodible bed, we observe how dune shape changes, eventually leading to the
elongation of isolated dunes with a different orientation. These experiments are
unique by their size and duration. Under natural conditions, they show that the
same wind regime can be associated with two dune growth mechanisms depending on
sand availability. The coexistence of these two dune growth mechanisms provides a
basis for examining the diversity of dune shapes on Earth or other planetary
bodies depending on local environmental conditions.

\section{Introduction}
\label{sec: intro}
Dunes record information about the environmental flows in which they formed and represent powerful
(paleo-)environmental proxies on Earth and other planetary bodies \cite{Ewin10a,Fent10,bLore,Lapo16,Lapo21,Chan22}.
However, given the variety of boundary and initial conditions to be considered, a major question
has been to assess the relative contribution of current and past wind regimes to the
diversity of dune size, shape and orientation \cite{Kocu05,Ewin10,Part09,Telf18,Lapo18}. To conduct
this inverse problem, there is still a lack of reliable experiments on eolian dune growth under
natural wind regimes. Notably, the elongation of isolated dunes on a nonerodible bed has never been
experimentally validated in the field and compared to the dynamics of periodic dune patterns in
areas of abundant sediment supply - a prerequisite to better understand the coexistence of dunes
with different orientations and the impact of seasonal wind reversals on eolian landscapes within
sand seas.

Sand availability recognized early as an important factor in dune morphodynamics \cite{bBagn,Wass83},
but its strong influence on dune growth mechanism has only recently been formalized by \citeA{Cour14}.
Past research mainly focused on the dune instability that arises over an unlimited layer of loose
sand \cite{Kenn63,Rich80,Andr02a,Gada19,Gada20a,Delo20}. Over short timescales, the linear regime of
this instability is responsible for the emergence of periodic
dunes, assuming that all modes (wavelengths) grow exponentially independently from each other
\cite{Lu21}. Over longer timescales, a non-linear regime of dune pattern coarsening dominated by
collisions and interactions between bedforms leads to an increase in dune amplitude and wavelength
in space and time \cite{Gao15b,Gada20a,Jarv22}. In areas of limited sediment supply, dunes adopt crescentic
shapes that can also elongate by depositing at their tips the sediment that is transported along
their crests \cite{Part09,Reff10,Zhan12,Cour14,Luca14,Gao15a,Luca15,Lu17,Casc18,Gada20}. These two
dune growth mechanisms rely differently on sand fluxes perpendicular to the crest. According to
dune-instability theory for unlimited sediment supplies, bedforms align along the direction for which
the sum of crest-normal fluxes reaches a maximum, i.e., the direction of highest growth rate
\cite{Rubi87,Ping14,Gada19}. Alternatively, dunes may elongate in the direction that allows
crest-normal fluxes from both sides of the crest to cancel each other out. This is the reason why
elongation requires at least bidirectional wind regimes and why, under specific multidirectional
regimes, star-dune arms can grow in various directions \cite{Zhan12}.

Using a set of landscape-scale experiments, we investigate here the formation and development of
dunes in different conditions of sand availability to asses the coexistence of these two dune growth
mechanisms. Collected data are analyzed to test under natural wind conditions the predictions of
\citeA{Cour14} regarding dune shape, orientations and dynamics (growth, migration or elongation), as
well as the associated sand fluxes at their crests. In addition, these data are used to quantify the
relationships between defect density, wavelength and amplitude of periodic dune patterns during the
coarsening phase in areas of unlimited sediment supply, and the characteristic length-scales
associated with the elongation phase of dunes lying on a non-erodible bed \cite{Rozi19}.


\begin{sidewaysfigure}
\includegraphics[width=53pc]{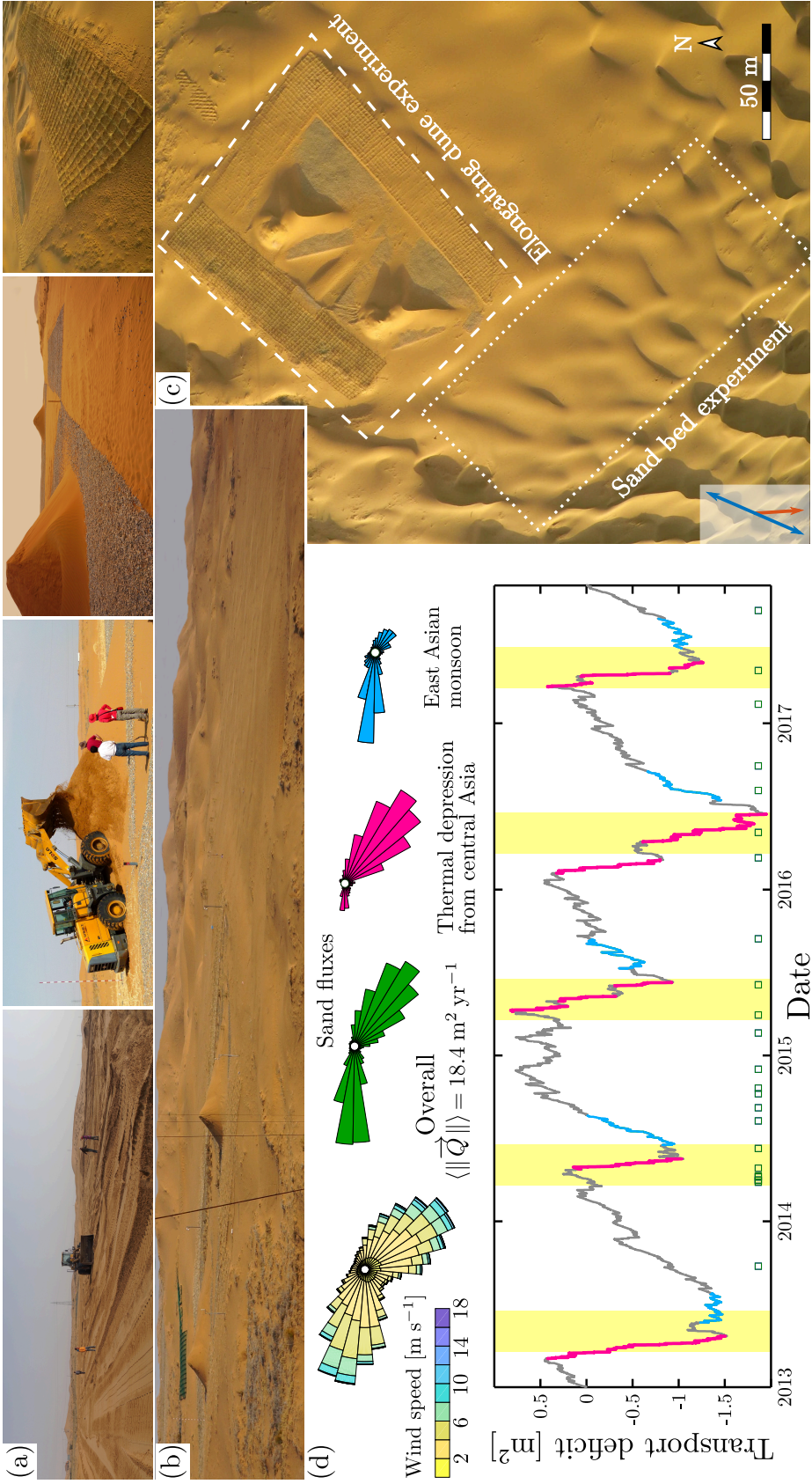}
\caption{
{\bf Landscape-scale experiments on dune growth in the Tengger desert}.
{\bf (a)}
Elongating dune experiment in October 2013 (37.560\textdegree{}~N, 105.033\textdegree{}~E).
After leveling (left), sand heaps are placed on a gravel bed (middle) surrounded by straw
checkerboards (right).
{\bf (b)}
Flat sand bed (on the right) and the elongating dune (on the left) experiments in April 2014.
{\bf (c)}
Aerial view of the experimental site in April 2015. Arrows show dune orientations
in the bed instability (blue) and the elongating modes (red) predicted from
wind data.
{\bf (d)}
Wind and sand flux roses from January 2013 to November 2017 and the corresponding
variations in transport deficit (Fig.~S4 and Supporting Text~3). Specific periods during
which primary and secondary winds are highlighted in red and blue, respectively (see the
corresponding flux roses). Yellow periods show springs. Squares indicate the dates of
topographic surveys.
}
\label{fig: fig1}
\end{sidewaysfigure}


\section*{Landscape-Scale Experiments}
Field experiments were continuously conducted from October 2013 to November 2017~in the Tengger Desert
at the southeastern edge of the Gobi basin in China, an area exposed to a bimodal wind regime \cite{Ping14}.
Two main experiments have been performed. The elongating dune experiment started in October 2013 by
placing two conical sand heaps $2.5$ and $3\,{\rm m}$ high on a flat gravel bed isolated from the
incoming sand by straw checkerboards (Fig.~1a). To the southwest, pre-existing dunes were also leveled
in April 2014 to form a flat rectangular bed $100\,{\rm m}$ long and $75\,{\rm m}$ wide (Figs.~1b,c).
The main axis of this sand bed experiment was aligned with the direction of the primary wind, which
blows from the northwest mainly in the spring when the Siberian high-pressure system weakens (Fig.~1d).
In summer, the easterly wind of the east-Asian monsoon dominates. This bimodal wind regime has a
divergence angle of 146\textdegree\ and a transport ratio of $1.5$, so that all conditions to observe
both growth mechanisms are met according to theory and numerical simulations \cite{Cour14,Gao15a}.
Sand heaps are located far enough from the sand bed experiment to ensure that the downstream flow field
does not affect dune dynamics in the zone of unlimited sediment supply, especially when the easterly
wind blows. Thus, the entire experimental setup was designed to promote dune elongation on the gravel
bed as well as dune growth and migration on the flat sand bed (arrows in Fig.~1c).

In this area of the Tengger desert, \citeA{Lu21} measured a mean grain size of $190\;\mu{\rm m}$ (Fig.~S1,
Supporting Text~1) and a threshold shear velocity of $0.23\pm0.04\;{\rm m\,s}^{-1}$ for aerodynamic
entrainment of sand grains. We use wind data from local meteorological towers and an airport located
$10\,{\rm km}$ east from the experimental dune field (Figs.~S2 and S3, Supporting Text~2). We calculate
the saturated sand flux on a flat sand bed and at the dune crests using the eolian transport law of
\citeA{Unga87} \cite<see also>[]{Dura11} and the formalism of \citeA{Cour14} (Tab.~S1, Supporting Text~3).
Using a reference system of concrete posts, more than 20 ground-based laser scans were performed to
frequently measure the evolution of surface elevation in the sand bed and the elongating dune experiments
over 4 years (see satellite images in Fig.~S5). Point density varied from 470 to 2,370 points per
${\rm m}^2$ with a centimeter height accuracy.


\begin{sidewaysfigure}
\includegraphics[width=53pc]{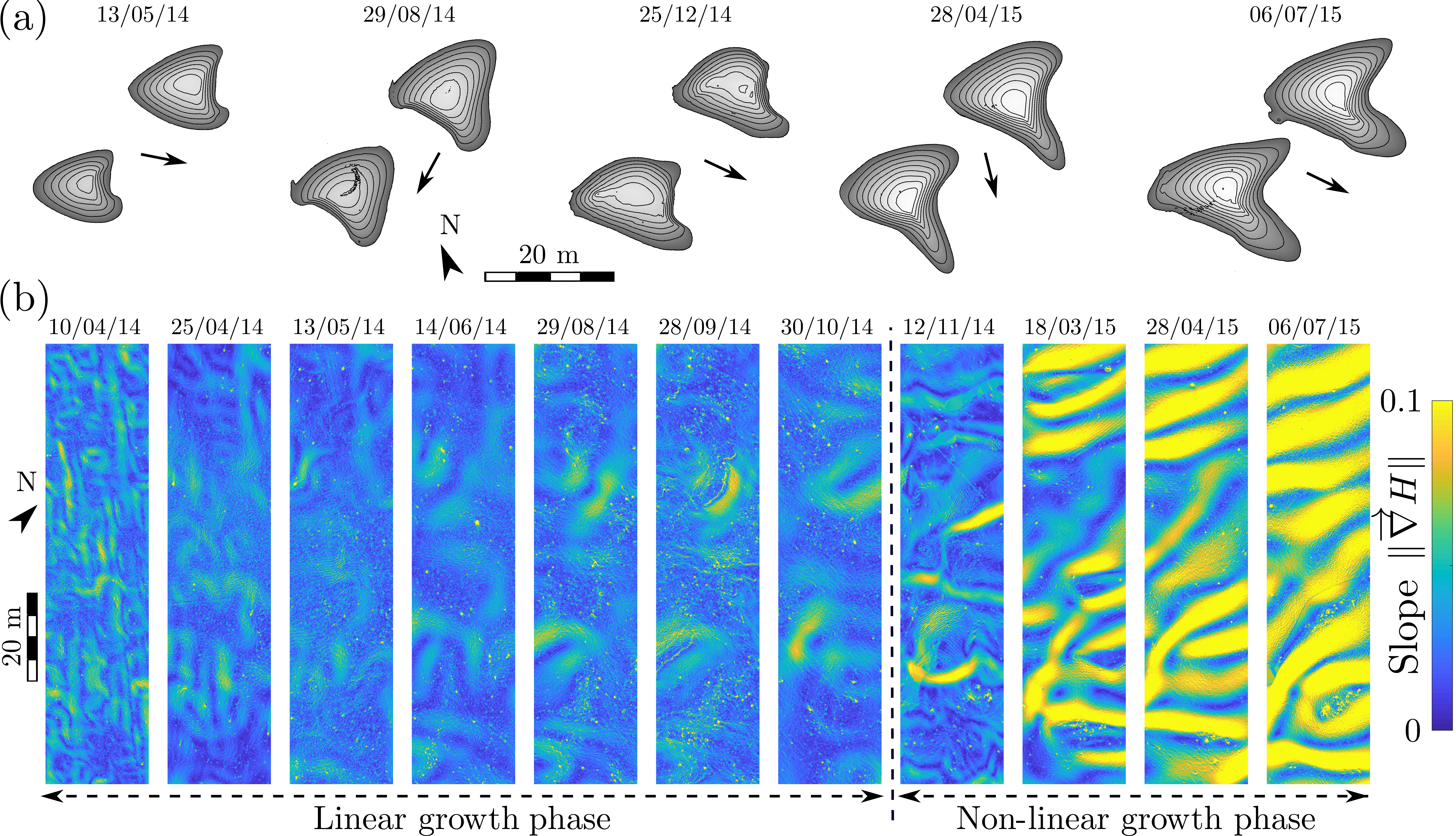}
\caption{
{\bf The early stages of dune growth from April 2014 to July 2015}.
{\bf (a)}
Evolution of the two sand heaps of the elongating dune experiments, from dome to
barchan and asymmetric barchan shapes. The difference in height between consecutive
contours is $25$~cm. The lowest contour corresponds to the elevation of the
nonerodible bed. Arrows indicate the resultant sand flux direction over the different
time intervals.
{\bf (b)}
Slope maps during incipient dune growth in the sand bed experiment. The colormap
is saturated to highlight the transition from the linear to the non-linear
phases of the bed instability. In non-saturated areas, ripples can be observed.
}
\label{fig: fig2}
\end{sidewaysfigure}


\section*{From Initial Dune Growth to Long-Term Dynamics}
The two initial sand heaps of the elongating dune experiment were not large enough ($33$ and $57\,{\rm m}^3$)
for dune instability or elongation to occur, and they rapidly took a dome dune shape under the
effect of the first wind reversals \cite{Bris04,Gao18}. Then, seasonal winds lead to complete reworking
of the dunes, resulting in successive reorientations of crescentic barchans until April 2015 (Fig. 2a).
To allow for dunes to increase their size, we added sand 10 times from May 2013 to September 2016,
always in the same places, to rebuild the original sand heaps (Fig.~S6, Supporting Text~4). This regular
supply is combined with incoming sand fluxes and exchanges of sediments between the two piles, which then
evolved into barchan dunes. These mass exchanges benefited to the dune to the southwest, which became
larger than the one to the northeast. As both dunes increased in size, their orientation stabilized,
adopting an asymmetric barchan shape with a southeast-facing slip face and a longer southern arm.

In the sand bed experiment, the linear phase of the dune instability under unlimited sediment supply was
observed from the earliest stage of dune growth, when the residual topography left by the leveling process
was smoothed and then disturbed by aeolian transport. During this initial phase, \citeA{Lu21} quantified
how growth rate varies with dune wavelength and showed that the highest growth rate was associated with an
intermediate length scale of approximately $15\,{\rm m}$. This most unstable mode (wavelength) eventually
emerged and prevailed across the entire experimental field in November 2014 for mean slope values of
$0.07$ ($\approx 4^{\rm o}$, for both lee and stoss slopes), when the amplitude of topography was less
than $20\,{\rm cm}$ (Fig. 2b).
The emergence of this periodic dune pattern reveals the predominance of a southwest-northeast bedform
alignment. Then, dunes continued migrating but started growing both in amplitude and wavelength as the
non-linear phase of the dune instability took over.

From October 2015 to November 2017, dune pattern coarsening in the sand bed experiment and elongation of
the largest asymmetric barchan in the isolated heaps experiment both occurred simultaneously. (Figs.~3a,
S7 and S8). In the sand bed experiment, dunes reached a height of $2\,{\rm m}$ and migrated at an average
speed of $5\;{\rm m}\,{\rm yr}^{-1}$ from June 2016 to June 2017 (Fig.~3b).
They formed a periodic pattern of oblique linear dunes at a 34\textdegree\ angle to the resultant sand
flux on a flat sand bed \cite{Hunt83,Ping14}. In the elongating dune experiment, a longitudinal dune
developed at an angle of less than 10 degrees to the resulting sand flux. Its total length increased by
more than $10\,{\rm m}$ from April 2016 to June 2017, with a height profile which decreased almost
linearly from the sand source area to the tip (Fig.~3c). These observations represent a direct validation
in the field of the elongation stages numerically evidenced by \citeA{Rozi19}.


\begin{sidewaysfigure}
\includegraphics[width=53pc]{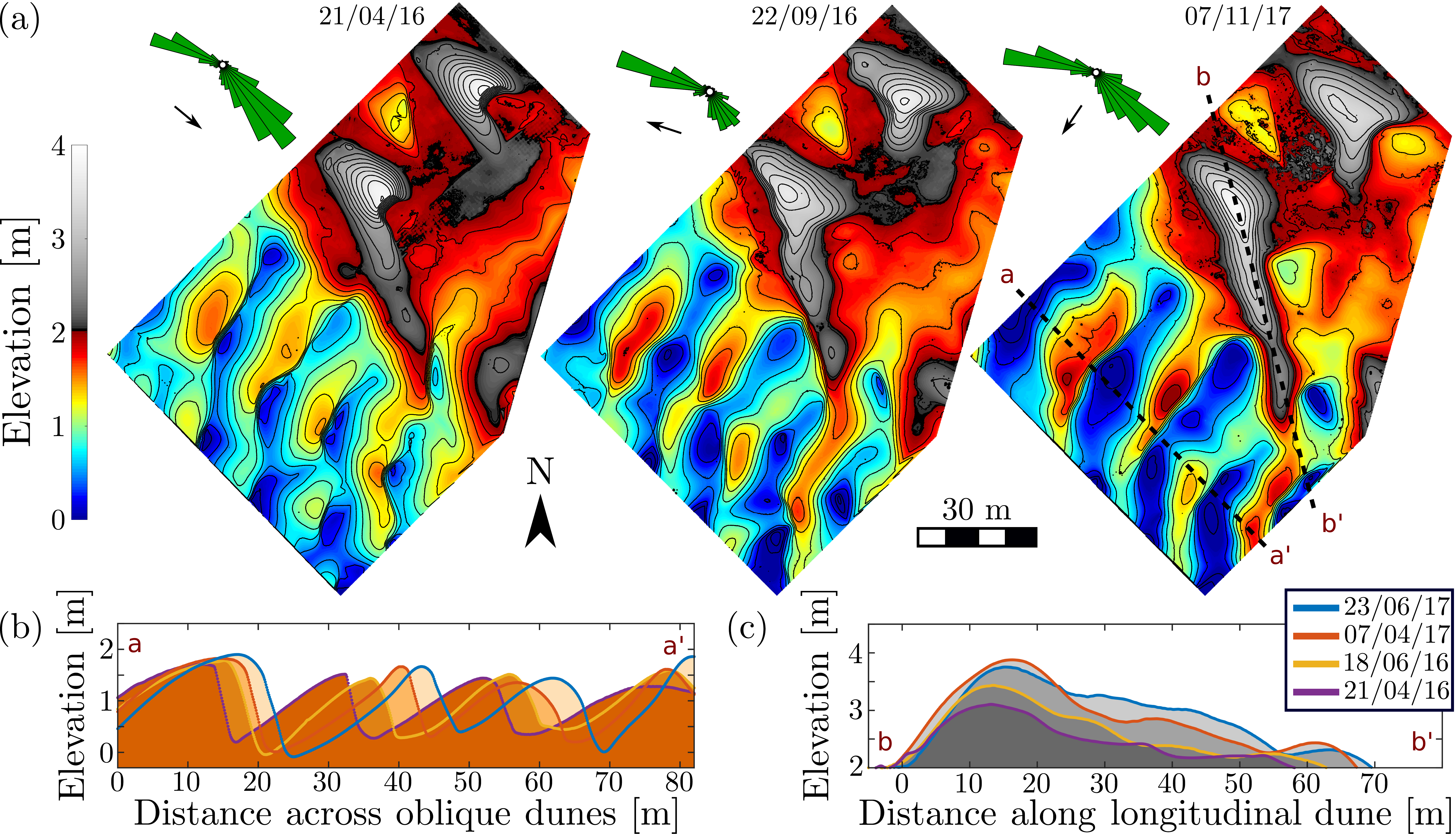}
\caption{{\bf Coexistence of two dune growth mechanisms starting in April 2016.}
{\bf (a)}
Surface elevation of the flat sand bed and the elongating dune experiments.
Flux roses show transport directionality over the different time intervals.
Arrows indicate the resultant sand flux direction.
{\bf (b)}
Migration of oblique linear dunes in the flat sand bed experiment.
{\bf (c)}
Elongation of the linear longitudinal dune to the southwest.
}
\label{fig: fig3}
\end{sidewaysfigure}


\section*{Quantification of Dune Morphodynamics}
Frequent  acquisition of topographic data allowed for a unique analysis of the two independent
growth mechanisms that occur simultaneously. Fig.~4a illustrates the coarsening dynamics of periodic
dune patterns and shows that the number of terminations (i.e, defect density) decreased as the
characteristic wavelengths of crests and troughs increase \cite{Day18}.
From April 2015 to November 2017, the wavelength increased from $15$ to $25$ meters, while the average
height $\langle h \rangle$ tripled from $0.5$ to $1.5$ meters. These variations lead to a continuous
increase in dune aspect ratio (height/wavelength) and an exponential relaxation towards an equilibrium
value (Fig.~4b). This coarsening dynamics was closely related to the southeastern migration of the dune
pattern, which was directly measured by cross-correlation between successive elevation maps and
predicted from the sand flux at the crest derived from wind data (Fig.~4c). As shown in Fig.~4d, dune
orientations of both the elongating dune and in the sand bed experiments were successively compared to
the model predictions of \citeA{Cour14}. The agreement between data-driven predictions and observed
orientations and dynamics validates the theoretical approach beyond the initial phase of the dune
instability \cite{Lu21}, under natural wind conditions and for interacting mature bedforms.

Using the elevation of the gravel bed as a reference level for the entire duration of the experiments,
we measure the dimensions and volume of the elongating dune as a function of time. Owing to episodic
sand input (Fig.~S6), the volume of the dune doubled from April 2014 to April 2015, while
keeping the aspect ratio between length and width constant (Fig.~4e). Then, a critical volume of
$150\,{\rm m}^3$ was reached above which the width of the dune stabilized as the southern arm elongated
by more than $10\,{\rm m}$ in less than a year (Fig.~4f). The positive correlation between the volume
and the length-to-width ratio of the dune continued through the end of 2017. Throughout this process,
the dune tip maintained a constant size with a transverse width of approximately $5\,{\rm m}$ and the
asymmetric barchan shape can be considered as a transient state in a continuous transition from a
barchan to an isolated linear dune shape.


\begin{figure}
\centerline{
\includegraphics[width=1.34\linewidth]{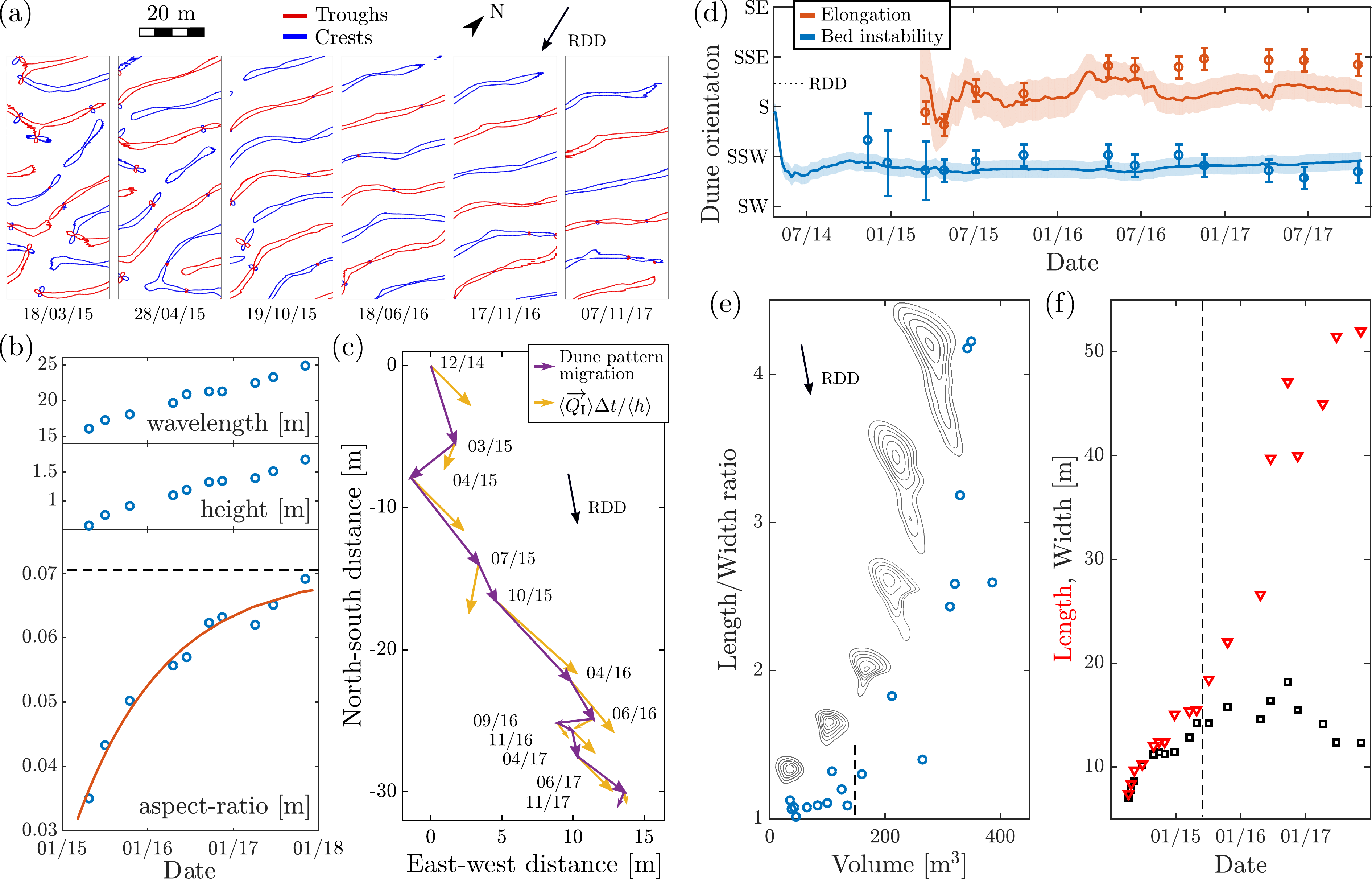}
}
\caption{
{\bf Dune morphodynamics}.
{\bf (a)}
Crests and troughs during dune pattern coarsening in the sand bed
experiment. Contours show isolines of
${\rm div}(\mbox{\boldmath$\nabla$} h/ \| \mbox{\boldmath$\nabla$} h \|)$.
Defects produce quadrupole-like structures. The black arrow shows the
resultant sand flux direction (RDD).
{\bf (b)}
Wavelength, mean amplitude, and aspect ratio of periodic dunes
in the sand bed experiment. The red line shows an exponential relaxation
towards a plateau value of 0.07 (i.e., a mean slope of 0.14 from troughs
to crests).
{\bf (c)}
Dune pattern migration obtained by cross-correlation between consecutive
elevation maps in the sand bed experiment. The predicted sand flux vectors
$\langle \overrightarrow{Q_{\rm I}}\rangle \Delta t/\langle h \rangle$
at the crest of these periodic dunes are computed from wind data over
the corresponding time interval (Tab.~S1 and Supporting Text~3).
{\bf (d)}
Orientations in the elongating (red) and sand bed (blue) experiments
measured from elevation maps by autocorrelation (dots) and predicted from
wind data over the cumulative time intervals for grain
sizes from $150$ to $250\,\mu{\rm m}$ and an aerodynamic roughness of
$10^{-4}$ to $10^{-2}\,{\rm m}$ (shaded areas). Errorbars show standard
deviations using different radii to integrate the angular energy distribution
in autocorrelograms.
{\bf (e)}
Length to width ratio of sand heaps in the elongating dune experiment with
respect to their volumes. Dunes are shown at different times using contour
lines.
{\bf (f)}
Sand heaps length and width with respect to time estimated from the
distribution of mass above the elevation of the nonerodible bed
(Supporting Text~4). Dashed lines show the onset of elongation.
}
\label{fig: fig4}
\end{figure}

\section*{Discussion}
In our landscape-scale experiments, we observed the development of different types of dunes depending
on sand availability, and quantified their orientation, dynamics (growth, migration or elongation) and
the corresponding sand fluxes at their crests. As shown in Fig.~4, these results not only directly
support the model predictions of \citeA{Cour14}, they also allow to document the early phases of aeolian
dune growth with an unprecedented resolution.

\citeA{Lu21} characterized the initial linear phase of the emergence of periodic dunes under unlimited
sediment supply from April to December 2014; here, we further characterized the non-linear phase of
pattern coarsening over more than 2 years. After the wavelength roughly doubled, the dune aspect-ratio
rapidly relaxes toward an equilibrium value, which is similar to that of the surrounding mature dunes
in the Tengger Desert \cite{Wen16}. Given the small dune size and their continuous growth, these
observations suggest that the dune aspect-ratio is mainly governed by the wind regime, including frequent
wind reversals and changes in wind speed, and not by a dune height saturation mechanism. Instead, these
changes in dune aspect-ratio certainly contributes to the final dune height by modifying the flow
properties over the dunes, especially the upwind shift between the wind speed and the bed topography
\cite{Clau13}. We also show that the overall migration of the dune pattern during the coarsening phase
can be predicted with reasonable accuracy using the normal-to-crest component of transport, with the
exception of time periods dominated by the secondary wind. Indeed, crest reversals are not taken into
account by the model; adding the strong increase in wind speed-up after wind reversals could
improve predictions \cite{Gao21}.

On a nonerodible bed, we experimentally validated the model for dune elongation and, in the process,
showed how larger and larger sand heaps successively produce dome, barchan, and asymmetric barchan shapes
\cite{Badd18}. A minimum dune volume of $150\;{\rm m}^3$ is required to integrate the local wind
variability and initiate elongation. While such threshold behavior has been numerically predicted by
\citeA[Fig.~1f]{Rozi19}, this is the first experimental evidence in the field of a causal relationship
between dune shape and size in multidirectional wind regimes. Supporting another prediction of
\citeA{Rozi19}, we verify that the minimum dune width at the tip is set by $\lambda_0\sin(\theta_1)$,
where $\lambda_0=9\;{\rm m}$ is the lower cutoff wavelength of the dune instability measured over the
same time period in the sand bed experiment \cite{Lu21}, and $\theta_1=38^{\rm o}$ is the angle
between the primary wind and the dune crestline. The stability of the elongating dune relies on the
balance between input and output fluxes occurring over the entire length of the dune according to wind
strength and orientation. As we regularly fed the original sand-heap and given the limited size of the
nonerodible bed, it is difficult to evaluate if a steady length was reached \cite{Rozi19}.
However, considering the resultant sand flux along the crest
$\langle\|\overrightarrow{Q_{\rm F}}\|\rangle=11.1\;{\rm m}^2\,{\rm yr}^{-1}$,
as derived from wind data (Tab.~S1, Supporting Text~3), we can compute the elongation rate,
$e=\langle\|\overrightarrow{Q_{\rm F}}\|\rangle/h_{\rm f}$,
where $h_{\rm f}$ is dune height at the dune front where width and height decrease rapidly. For a dune
front of meter-scale height, this estimate agrees with the elongation rate of
$e\approx 10\;{\rm m}\,{\rm yr}^{-1}$
observed from April 2016 to June 2017 (see Fig.~3c).

Our landscape-scale experiments conducted in an active dune field provide a quantitative
characterization of dune morphogenesis under the natural action of wind for different conditions of
sand availability. The elongation of a longitudinal linear dune from a fixed source of sand placed on
a nonerodible bed is synchronous with the emergence and the coarsening of periodic oblique dunes in an
area of unlimited sediment supply. This demonstrates that two independent dune growth mechanisms coexist
and regulate dune shape, orientation and dynamics according to the boundary conditions and the nature
of the bed (i.e., an erodible sand bed or a nonerodible ground). Based on the experimental design and
observed bedforms, we show that all these dune properties can be predicted from wind data with
increasing accuracy by models that simulate not only the initial growth of dunes \cite{Gada20a,Lu21}
but also their morphodynamic response to specific wind cycles \cite{Cour14,Gao18,Gao21}.

The coexistence of two growth mechanisms naturally explains some of the natural complexities of dune
fields, and offer new perspectives to Earth-based and planetary geologists studying the evolution of
sand seas of the solar system. Dunes and superimposed bedforms with different shapes and orientations
can now be studied according to elementary dune types relying on two independent dune growth
mechanisms. This effort devoted to questions of pattern emergence and organization is critical to
predict the evolution of eolian dune systems according to changes in climate and wind properties.

\newpage


\section*{Data Availability}
\noindent
All study data are included in the article and/or SI Appendix. Elevation and wind data are available
on https://doi.org/10.6084/m9.figshare.17817494.v4.


\section*{Author contributions}
\noindent
P.L. and C.N. carried out all statistical data analysis and numerical simulations, helped by C.G..
P.L., C.N., P.C., S.R. and S.C.P. designed the experimental study. P.L., C.N. and Z.D. managed the
experimental site and P.L. led the acquisition of data in the field. Z.A. performed all laser scans.
C.N., P.C., S.R. and S.C.P. wrote the manuscript. All authors participate to data acquisition and
discussed the results.


\acknowledgments
We acknowledge financial support from National
Natural Science Foundation of China Grants 41871011 and 41930641,
Laboratoire d’Excellence UnivEarthS Grant ANR-10-LABX-0023, Initiative
d’Excellence Universit\'e de Paris Grant ANR-18-IDEX-0001, French
National Research Agency Grants ANR-12-BS05-001-03/EXO-DUNES and
ANR-17-CE01-0014/SONO, National Science Center of Poland Grant
2016/23/B/ST10/01700, and the French Chinese International Laboratory on
Sediment Transport and Landscape Dynamics.

\newpage


\renewcommand\refname{References From the Supporting Information}

\end{document}